\documentclass[aps,prl,superscriptaddress,showpacs,showkeys,floatfix,twocolumn]{revtex4}

\usepackage{graphicx}	

\usepackage{dcolumn}	

\newcommand{\AN}{\mbox{$A_N$}}
\newcommand{\ANsub}{\mbox{$A_N^{\pi^0}$}}
\newcommand{\ANtot}{\mbox{$A_N^{\rm peak}$}}
\newcommand{\ANbg}{\mbox{$A_N^{\rm bg}$}}
\newcommand{\ANneg}{\mbox{$A_N^{h^-}$}}
\newcommand{\ANpos}{\mbox{$A_N^{h^+}$}}
\newcommand{\pperp}{\mbox{$p_T$}}

\begin{document}

\title{Measurement of Transverse Single-Spin Asymmetries for Mid-rapidity Production 
of Neutral Pions and Charged Hadrons in Polarized p+p Collisions at $\sqrt{s}=200$~GeV}

\newcommand{\abilene}{Abilene Christian University, Abilene, TX 79699, USA}
\newcommand{\acadsin}{Institute of Physics, Academia Sinica, Taipei 11529, Taiwan}
\newcommand{\banaras}{Department of Physics, Banaras Hindu University, Varanasi 221005, India}
\newcommand{\barc}{Bhabha Atomic Research Centre, Bombay 400 085, India}
\newcommand{\bnl}{Brookhaven National Laboratory, Upton, NY 11973-5000, USA}
\newcommand{\caucr}{University of California - Riverside, Riverside, CA 92521, USA}
\newcommand{\ciae}{China Institute of Atomic Energy (CIAE), Beijing, People's Republic of China}
\newcommand{\cns}{Center for Nuclear Study, Graduate School of Science, University of Tokyo, 7-3-1 Hongo, Bunkyo, Tokyo 113-0033, Japan}
\newcommand{\columbia}{Columbia University, New York, NY 10027 and Nevis Laboratories, Irvington, NY 10533, USA}
\newcommand{\dapnia}{Dapnia, CEA Saclay, F-91191, Gif-sur-Yvette, France}
\newcommand{\debrecen}{Debrecen University, H-4010 Debrecen, Egyetem t{\'e}r 1, Hungary}
\newcommand{\fsu}{Florida State University, Tallahassee, FL 32306, USA}
\newcommand{\gsu}{Georgia State University, Atlanta, GA 30303, USA}
\newcommand{\hiroshima}{Hiroshima University, Kagamiyama, Higashi-Hiroshima 739-8526, Japan}
\newcommand{\ihepprot}{IHEP Protvino, State Research Center of Russian Federation, Institute for High Energy Physics, Protvino, 142281, Russia}
\newcommand{\isu}{Iowa State University, Ames, IA 50011, USA}
\newcommand{\jinrdubna}{Joint Institute for Nuclear Research, 141980 Dubna, Moscow Region, Russia}
\newcommand{\kaeri}{KAERI, Cyclotron Application Laboratory, Seoul, South Korea}
\newcommand{\kangnung}{Kangnung National University, Kangnung 210-702, South Korea}
\newcommand{\kek}{KEK, High Energy Accelerator Research Organization, Tsukuba, Ibaraki 305-0801, Japan}
\newcommand{\kfki}{KFKI Research Institute for Particle and Nuclear Physics of the Hungarian Academy of Sciences (MTA KFKI RMKI), H-1525 Budapest 114, POBox 49, Budapest, Hungary}
\newcommand{\korea}{Korea University, Seoul, 136-701, Korea}
\newcommand{\kurchatov}{Russian Research Center ``Kurchatov Institute", Moscow, Russia}
\newcommand{\kyoto}{Kyoto University, Kyoto 606-8502, Japan}
\newcommand{\labllr}{Laboratoire Leprince-Ringuet, Ecole Polytechnique, CNRS-IN2P3, Route de Saclay, F-91128, Palaiseau, France}
\newcommand{\lawllnl}{Lawrence Livermore National Laboratory, Livermore, CA 94550, USA}
\newcommand{\losalamos}{Los Alamos National Laboratory, Los Alamos, NM 87545, USA}
\newcommand{\lpc}{LPC, Universit{\'e} Blaise Pascal, CNRS-IN2P3, Clermont-Fd, 63177 Aubiere Cedex, France}
\newcommand{\lund}{Department of Physics, Lund University, Box 118, SE-221 00 Lund, Sweden}
\newcommand{\muenster}{Institut f\"ur Kernphysik, University of Muenster, D-48149 Muenster, Germany}
\newcommand{\myongji}{Myongji University, Yongin, Kyonggido 449-728, Korea}
\newcommand{\nagasaki}{Nagasaki Institute of Applied Science, Nagasaki-shi, Nagasaki 851-0193, Japan}
\newcommand{\newmex}{University of New Mexico, Albuquerque, NM 87131, USA}
\newcommand{\nmsu}{New Mexico State University, Las Cruces, NM 88003, USA}
\newcommand{\ornl}{Oak Ridge National Laboratory, Oak Ridge, TN 37831, USA}
\newcommand{\orsay}{IPN-Orsay, Universite Paris Sud, CNRS-IN2P3, BP1, F-91406, Orsay, France}
\newcommand{\pnpi}{PNPI, Petersburg Nuclear Physics Institute, Gatchina, Russia}
\newcommand{\riken}{RIKEN, The Institute of Physical and Chemical Research, Wako, Saitama 351-0198, Japan}
\newcommand{\rikjrbrc}{RIKEN BNL Research Center, Brookhaven National Laboratory, Upton, NY 11973-5000, USA}
\newcommand{\saispbstu}{Saint Petersburg State Polytechnic University, St. Petersburg, Russia}
\newcommand{\saopaulo}{Universidade de S{\~a}o Paulo, Instituto de F\'{\i}sica, Caixa Postal 66318, S{\~a}o Paulo CEP05315-970, Brazil}
\newcommand{\seoulnat}{System Electronics Laboratory, Seoul National University, Seoul, South Korea}
\newcommand{\stonybrkc}{Chemistry Department, Stony Brook University, SUNY, Stony Brook, NY 11794-3400, USA}
\newcommand{\stonycrkp}{Department of Physics and Astronomy, Stony Brook University, SUNY, Stony Brook, NY 11794, USA}
\newcommand{\subatech}{SUBATECH (Ecole des Mines de Nantes, CNRS-IN2P3, Universit{\'e} de Nantes) BP 20722 - 44307, Nantes, France}
\newcommand{\tenn}{University of Tennessee, Knoxville, TN 37996, USA}
\newcommand{\titech}{Department of Physics, Tokyo Institute of Technology, Tokyo, 152-8551, Japan}
\newcommand{\tsukuba}{Institute of Physics, University of Tsukuba, Tsukuba, Ibaraki 305, Japan}
\newcommand{\vandy}{Vanderbilt University, Nashville, TN 37235, USA}
\newcommand{\waseda}{Waseda University, Advanced Research Institute for Science and Engineering, 17 Kikui-cho, Shinjuku-ku, Tokyo 162-0044, Japan}
\newcommand{\weizmann}{Weizmann Institute, Rehovot 76100, Israel}
\newcommand{\yonsei}{Yonsei University, IPAP, Seoul 120-749, Korea}
\affiliation{\abilene}
\affiliation{\acadsin}
\affiliation{\banaras}
\affiliation{\barc}
\affiliation{\bnl}
\affiliation{\caucr}
\affiliation{\ciae}
\affiliation{\cns}
\affiliation{\columbia}
\affiliation{\dapnia}
\affiliation{\debrecen}
\affiliation{\fsu}
\affiliation{\gsu}
\affiliation{\hiroshima}
\affiliation{\ihepprot}
\affiliation{\isu}
\affiliation{\jinrdubna}
\affiliation{\kaeri}
\affiliation{\kangnung}
\affiliation{\kek}
\affiliation{\kfki}
\affiliation{\korea}
\affiliation{\kurchatov}
\affiliation{\kyoto}
\affiliation{\labllr}
\affiliation{\lawllnl}
\affiliation{\losalamos}
\affiliation{\lpc}
\affiliation{\lund}
\affiliation{\muenster}
\affiliation{\myongji}
\affiliation{\nagasaki}
\affiliation{\newmex}
\affiliation{\nmsu}
\affiliation{\ornl}
\affiliation{\orsay}
\affiliation{\pnpi}
\affiliation{\riken}
\affiliation{\rikjrbrc}
\affiliation{\saispbstu}
\affiliation{\saopaulo}
\affiliation{\seoulnat}
\affiliation{\stonybrkc}
\affiliation{\stonycrkp}
\affiliation{\subatech}
\affiliation{\tenn}
\affiliation{\titech}
\affiliation{\tsukuba}
\affiliation{\vandy}
\affiliation{\waseda}
\affiliation{\weizmann}
\affiliation{\yonsei}
\author{S.S.~Adler}	\affiliation{\bnl}
\author{S.~Afanasiev}	\affiliation{\jinrdubna}
\author{C.~Aidala}	\affiliation{\bnl}
\author{N.N.~Ajitanand}	\affiliation{\stonybrkc}
\author{Y.~Akiba}	\affiliation{\kek} \affiliation{\riken}
\author{J.~Alexander}	\affiliation{\stonybrkc}
\author{R.~Amirikas}	\affiliation{\fsu}
\author{L.~Aphecetche}	\affiliation{\subatech}
\author{S.H.~Aronson}	\affiliation{\bnl}
\author{R.~Averbeck}	\affiliation{\stonycrkp}
% ...
\collaboration{PHENIX Collaboration} \noaffiliation

\date{\today}

\begin{abstract}
The transverse single-spin asymmetries of neutral pions and non-identified charged hadrons 
have been measured at mid-rapidity in polarized proton-proton collisions at $\sqrt{s} = 200$~GeV.  The 
data cover a transverse momentum (\pperp) range 0.5-5.0 GeV/$c$ for charged hadrons 
and 1.0-5.0 GeV/$c$ for neutral pions, at a Feynman-$x$ ($x_F$) value of approximately zero.  
The asymmetries seen in this previously unexplored kinematic region are consistent with zero 
within statistical errors of a few percent. 
In addition, the inclusive charged hadron 
cross section at mid-rapidity from $0.5 < \pperp < 7.0$ GeV/$c$ is presented and compared to 
NLO pQCD calculations.  Successful description of the unpolarized cross section above 
$\sim 2$~GeV/$c$ using NLO pQCD 
suggests that pQCD is applicable in the interpretation of the asymmetry results 
in the relevant kinematic range.
\end{abstract}



\maketitle


The measurement of transverse single-spin asymmetries (SSAs) in proton-proton collisions 
and deep-inelastic lepton-nucleon scattering (DIS) probes the 
quark and gluon structure of transversely 
polarized nucleons.  Large transverse SSAs have been observed in a
number of spin-dependent proton-proton scattering experiments at energies ranging from 
$\sqrt{s} = 5 - 10$~GeV. Asymmetries approaching 30\% were observed in inclusive pion 
production at transverse momentum (\pperp) 
up to 1.2~GeV/$c$ and Feynman-$x$ ($x_F$) up to 0.8 \cite{Dragoset:1978gg,Allgower:2002qi}. 
At mid-rapidity
and $x_T = \frac{2p_T}{\sqrt{s}}$ up to 0.8, asymmetries were also observed in inclusive $\pi^0$ and $\pi^+$
production but not in $\pi^-$ production \cite{Antille:1980th,Saroff:1989gn,Apokin:1990ik}. 
At higher center-of-mass energies of 20 and 200~GeV, $\pi^+$, $\pi^-$, and $\pi^0$ 
asymmetries were found to persist at large $x_F$
 \cite{Adams:1991rw,Adams:1991cs,Adams:2003fx} while the asymmetry 
in $\pi^0$ production at mid-rapidity was found to be consistent with zero at $\sqrt{s} = 20$~GeV and for 
$\pperp< 4$~GeV/$c$ \cite{Adams:1994yu}. Non-zero transverse 
asymmetries have also been observed in semi-inclusive DIS experiments 
\cite{Airapetian:1999tv,Airapetian:2001iy,Airapetian:2004tw}.

Three different mechanisms have been studied as the possible origin of transverse SSAs 
in hadron collisions at high energies:  (1) Transversity distributions, the quark spin distributions 
in a transversely polarized proton, can give rise to SSAs in combination with spin-dependent fragmentation 
functions (FFs), e.g. the Collins function \cite{Collins:1992kk}.  Spin-dependent FFs serve as 
analyzers for the transverse spin of the struck quark.  
(2) Quark and gluon distributions that are asymmetric in the transverse intrinsic parton 
momentum, $k_T$, first suggested by Sivers \cite{Sivers:1989cc}, 
can lead to SSAs.  (3) Alternatively, interference between quark and gluon fields in the 
initial or final state can also generate SSAs \cite{Qiu:1998ia,Kanazawa:2000hz}.  Sivers 
parton distributions can exist both for quarks and gluons, and a  
possible connection to orbital angular momentum of partons in the nucleon 
has been suggested \cite{Sivers:1989cc,Burkardt:2003je}.  

It is expected that SSAs measured at the Relativistic Heavy Ion Collider (RHIC) result from 
a combination of these three
effects (see \cite{Adams:2003fx} and references therein).  Model calculations leading to 
predictions for the Sivers and transversity distributions have been performed to describe 
existing data at forward rapidities.
Precision measurements of SSAs
in different regions of $x_F$ and \pperp\ and their QCD analysis may serve to 
quantify contributions
from the competing mechanisms. In this Letter we present first measurements
of transverse single-spin asymmetries at mid-rapidity and collider energies.

These data were collected during the 2001-2 
polarized proton run at RHIC, in which approximately 0.15 $pb^{-1}$ of integrated luminosity 
were collected using the PHENIX detector.  
Two beams of 55 bunches of polarized protons, with approximately 
$5\times10^{10}$ protons per bunch, were injected into 
RHIC and accelerated to 100 GeV each.  

Measurements of the unpolarized production of charged hadrons and of the spin-dependent 
production of both neutral pions and charged hadrons were made in the central arms of 
the PHENIX detector.  These cover a pseudorapidity 
range of $|\eta| < 0.35$ and two azimuthal angle intervals of $\Delta\phi= 90^\circ$, offset 
$33.75^\circ$ from vertical
 \cite{Adcox:2003zm}.  

A minimum-bias (MB) collision trigger and the vertex position in the beam direction 
are provided by two beam-beam counters (BBCs) \cite{Allen:2003zt}.  The BBCs, which cover 2$\pi$ in azimuth and 
$3.0 < |\eta| < 3.9$, are sensitive to charged particles and select approximately 
half of the total inelastic proton-proton cross section.  A $\pm 30$~cm event vertex cut was applied 
for all analyses, corresponding to the central arm acceptance.  The approximate vertex resolution was 2~cm in the beam direction.  

Charged-particle tracks from MB events were reconstructed using a drift chamber and pad chambers 
\cite{Adcox:2003zp} as well as the collision vertex, which is the assumed point of origin because the tracking chambers are placed 
outside the magnetic field.  Thus charged particles that do not originate at the vertex have incorrectly reconstructed 
momentum, leading to low-momentum, long-lived particle decays (e.g. $K^{\pm}$, $K^0_L$) and conversion electrons 
as the two main sources of background.

For the charged hadron cross section, approximately 17 million MB events were analyzed.  
The luminosity was measured as $N_{\rm BBC}/\sigma_{\rm BBC}$ with $\sigma_{\rm BBC} = 21.8$~mb 
\cite{Adler:2003pb}, accounting for the fraction 
of the yield for which the MB condition was satisfied.  Backgrounds were estimated and 
subtracted statistically following the method of \cite{Adler:2003au}: conversion electrons were estimated using the different 
response of the ring-imaging $\check{\rm C}$erenkov detector (RICH)
\cite{Aizawa:2003zq} to electrons and charged pions, and decay particles were estimated using the track bend in 
the residual magnetic field in the tracking detectors.  Weak decays of short-lived particles, mainly $K^0_S$, $\Lambda$, 
and $\bar{\Lambda}$, remain, especially when they decay close to the vertex.  Based on a Monte Carlo simulation, 
the reported cross section was reduced by 7\% over the entire \pperp\ range to correct for these decays.  

\begin{figure}[thb]
\includegraphics[width=1.0\linewidth]{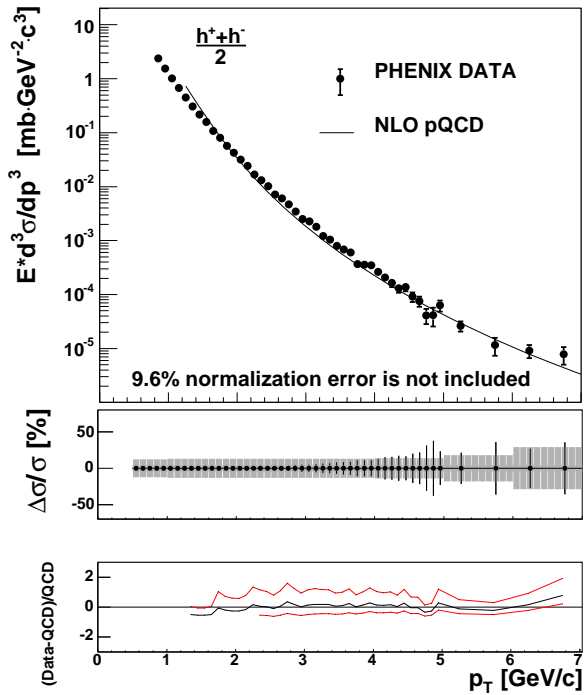}
\caption{\label{fig:chgcs}
Top panel: Invariant cross section vs. \pperp\ for the production of
charged hadrons at mid-rapidity, averaged over sign.  A 9.6\%
normalization uncertainty is not shown.  The curve represents an NLO pQCD
calculation at a renormalization scale of \pperp\
\protect\cite{Jager:2002xm}.  Middle panel: The relative statistical
(points) and point-to-point systematic (band) errors.  Bottom panel: The
relative difference between the data and the theory with scales of
\pperp/2 (lower curve), \pperp, and 2\pperp\ (upper curve).
}
\end{figure}

The unpolarized cross 
section for inclusive charged hadron production at mid-rapidity is presented 
in Fig.~\ref{fig:chgcs} and Table~\ref{tbl:ch_spectra}.  The dominant systematic uncertainty 
for $\pperp > 5$ GeV/$c$ is from the background subtraction, while for $\pperp < 5$ GeV/$c$ 
it is due to the weak-decay correction.  There is a 9.6\% normalization uncertainty due to 
the luminosity measurement. In Fig.~\ref{fig:chgcs} the cross section 
is compared to a next-to-leading-order (NLO) pQCD calculation using the CTEQ6M \cite{Pumplin:2002vw} 
parton distribution functions and Kniel-Kramer-P\"{o}tter FFs \cite{Kniehl:2000hk} and found to be consistent 
above $\pperp \sim 2$ GeV/$c$.  
The unpolarized cross sections for mid-rapidity and forward production 
of neutral pions have also been measured in 200-GeV 
proton-proton collisions at RHIC \cite{Adler:2003pb,Adams:2003fx} and have been found to agree well with NLO pQCD calculations
 \cite{Aversa:1988vb,deFlorian:2002az,Jager:2002xm}.  
The agreement between all 
of these unpolarized measurements and the theoretical calculations
indicates that NLO pQCD is applicable in interpreting polarized data at $\sqrt{s} = 200$~GeV 
and provides a solid theoretical foundation for the study of the spin structure of the proton at RHIC.

\begin{table}[ht]
\caption{\label{tbl:ch_spectra}
Selected invariant cross section values for $(h^++h^-)/2$ corresponding to
Fig.~\ref{fig:chgcs}.  A 9.6\% normalization uncertainty is not included.}
\begin{ruledtabular} \begin{tabular}{cccc}
               &     inv. cross     & stat.     & syst.     \\
$\rm{p_T}$     &       section      & error     & error     \\
$[\rm{GeV}/\it{c}]$         &     [$\rm{mb/GeV^2}$]      &     [\%]     &     [\%]     \\
\hline
0.55 & $1.06\times10^1$    &  0.2 &  8 \\
1.05 & $1.02\times10^0$    &  0.4 &  8.5 \\
1.55 & $1.58\times10^{-1}$ &  0.9 &  8.5 \\
2.05 & $3.16\times10^{-2}$ &  1.7 &  9 \\
2.55 & $7.14\times10^{-3}$ &  3.2 &  9 \\
3.05 & $2.28\times10^{-3}$ &  5.1 &  9 \\
3.55 & $6.86\times10^{-4}$ &  8.5 &  9 \\
4.05 & $2.63\times10^{-4}$ & 12.8 &  9.5 \\
4.55 & $9.12\times10^{-5}$ & 20.2 &  9.5 \\
5.25 & $2.61\times10^{-5}$ & 21.2 &  16 \\
5.75 & $1.16\times10^{-5}$ & 36.3 &  16 \\
6.25 & $9.12\times10^{-6}$ & 27.2 &  28 \\
6.75 & $7.82\times10^{-6}$ & 35.7 &  28 \\
\end{tabular} \end{ruledtabular}
\end{table}

The stable spin direction of the protons through 
acceleration and storage is vertical, and there is an approximately equal number of bunches filled
with the spin of the protons up as there is down..  With both beams polarized, single-spin 
analyses were performed by taking into account the spin states of one beam, averaging 
over those of the other.  The beam polarization at 100 GeV was obtained using the same analyzing power
($A_N^{pC}$) in proton-carbon elastic scattering in the Coulomb-nuclear interference
region measured at 22 GeV (see \cite{Jinnouchi:2003cp} and references therein), near RHIC injection
energy. The average beam polarization was 15$\pm$5\%.  

The left-right transverse single-spin asymmetry, \AN, can be extracted using 
\begin{equation}
A_N = \frac{1}{P_{\rm b}} \left( \frac{\sigma^{\uparrow} -
  \sigma^{\downarrow}}{\sigma^{\uparrow} + \sigma^{\downarrow}} \right)
 = \frac{1}{P_{\rm b}} \left(  \frac{N^{\uparrow} -
  \mathcal{R}N^{\downarrow}}{N^{\uparrow} + \mathcal{R}N^{\downarrow}} \right),
\label{eq:lumiformula}
\end{equation}
where $P_{\rm b}$ is the beam polarization, $\sigma^{\uparrow}$ ($\sigma^{\downarrow}$) the 
production cross section when the protons in the bunch are polarized up (down), $N^{\uparrow}$ 
($N^{\downarrow}$) the experimental yield from up- (down-) polarized bunches, and 
$\mathcal{R}=\mathcal{L}^{\uparrow}/\mathcal{L}^{\downarrow}$ the relative integrated luminosity of bunches of opposite polarization sign.  
The above formula as written applies to yields observed to the left of the polarized beam.  
An overall minus sign is required for yields observed to the right of the polarized beam.
Alternatively, we derive the asymmetry using 
\begin{equation}
\AN = \frac{1}{P_{\rm b}} \left(  \frac{\sqrt{N^{\uparrow}_{L}N^{\downarrow}_{R}} -
  \sqrt{N^{\downarrow}_{L}N^{\uparrow}_{R}}}{\sqrt{N^{\uparrow}_{L}N^{\downarrow}_{R}} +
  \sqrt{N^{\downarrow}_{L}N^{\uparrow}_{R}}} \right),
\label{eq:sqrtformula}
\end{equation}
which calculates a single value for the asymmetry taking into account yields from both the 
left ($N_L$) and right ($N_R$) sides of the polarized beam and provides a consistency check on the relative 
luminosity \cite{Spinka:1999vv}.

The BBCs were used to determine the relative luminosity ($\mathcal{R}$ in Eq.~\ref{eq:lumiformula}) 
between bunches of opposite polarization sign fill-by-fill.  
A typical $\mathcal{R}$ for the data sample analyzed here 
was approximately 1.09, measured to better than $10^{-3}$.
In the asymmetry analysis of charged hadrons, which utilized $\sim13$M minimum-bias events,
it was required that there be no hits in the RICH in order to eliminate electrons from 
photon conversions which mimic high-\pperp\ charged tracks.  The momentum threshold for production 
of $\check{\rm C}$erenkov radiation 
by pions was 4.7~GeV/$c$, allowing the RICH veto to preserve nearly all charged pions.  
The electron contamination in the final data sample was 
less than 1\%.  The decay background from long-lived particles was 
less than 5\%.  

Neutral pions were reconstructed via their decay to two photons using finely segmented 
($\Delta\phi \times \Delta\eta \approx 0.01 \times 0.01$) electromagnetic calorimeters 
(EMCal) \cite{Aphecetche:2003zr}.  Photon clusters were selected by their shower shape and 
a charged track veto.  Approximately 18M events recorded by an 
EMCal-based high-energy photon trigger in coincidence with the BBC collision trigger 
were analyzed \cite{Adler:2003pb}.   The trigger efficiency for neutral 
pions varied from $\sim24$\% in the 1-2~GeV/$c$ bin to $\sim78$\% in the 4-5~GeV/$c$ bin.  
Only triggered events were used in this analysis.  The 
$\pi^0$ peak widths varied from 13.2 MeV/$c^2$ in the 1-2~GeV/$c$ bin to 10.6 MeV/$c^2$ in the 4-5~GeV/$c$ 
bin.  The contribution from combinatorial background ranged from 34\% to 5\% across these bins; 
in order to avoid errors associated with peak extraction it was not subtracted.  

The asymmetry for neutral pions and charged hadrons was determined for each fill using Eq.~\ref{eq:lumiformula}, 
then averaged over all fills.  The contribution to the $\pi^0$ asymmetry by the background under the peak was 
estimated by calculating the asymmetry of 50-MeV/$c^2$ regions on both sides of the signal, from 
60-110~MeV/$c^2$ and 170-220~MeV/$c^2$ (see Table~\ref{tbl:anpiz}).  The asymmetry of the 
signal region and its uncertainty were then corrected using
\begin{equation}
A_N^{\pi^0}  =  \frac{A_N^{\rm peak} - r\ANbg}{1-r}, 
\sigma_{A_N^{\pi^0}}  = 
\frac{\sqrt{\sigma_{A_N^{\rm peak}}^{2} +
    r^{2}\sigma_{A_N^{\rm bg}}^{2}}}{1-r}
\end{equation}
where $r$ is the fraction of background under the peak.
\begin{figure}[bt]
\includegraphics[width=1.0\linewidth]{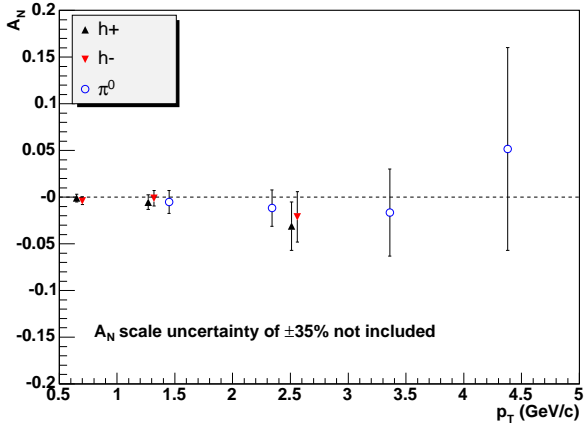}
\caption{\label{fig:an}
Mid-rapidity neutral pion and charged hadron transverse single-spin
asymmetry, \AN, vs. transverse momentum. Points for positive hadrons have
been shifted down by 50 MeV/$c$ to improve readability. The error bars
indicate statistical uncertainties.  
} 
\end{figure}

As the dominant systematic uncertainty is expected to be from the determination of the relative luminosity, 
systematic errors were evaluated by direct comparison of the asymmetry values 
calculated using Eq.~\ref{eq:lumiformula} and Eq.~\ref{eq:sqrtformula}.  Any potential effect should be 
the same for both the charged hadron and neutral pion analyses.  No \pperp\ dependence was expected or 
observed; therefore, we take the weighted average of the systematic uncertainties calculated for each bin, 0.002, 
as the overall, uniform systematic uncertainty.

The resulting asymmetries are plotted vs. \pperp\ in Fig.~\ref{fig:an} and
shown in Tables~\ref{tbl:anpiz} and \ref{tbl:anchg}.
The asymmetries are consistent with 
zero over the entire transverse momentum range.  
\begin{table}
\caption{\label{tbl:anpiz}
Neutral pion transverse single-spin asymmetry values and statistical
uncertainties for all photon pairs falling within the $\pi^0$ mass peak,
for the background ($\rm bg$), and for the $\pi^0$ background-corrected.  
The third column ($r$) indicates the background contribution under the
$\pi^0$ peak.  
 An \AN\ scale uncertainty of $\pm$35\% is not included.
}
\begin{ruledtabular} \begin{tabular}{cccccc}
\pperp\   & $\langle \pperp \rangle$  & $r$  &\ANtot\        & \ANbg\        & \ANsub\       \\ 
(GeV/$c$) & (GeV/$c$)                 & (\%) & (\%)          & (\%)          & (\%)          \\ 
\hline 
1-2       & 1.45                      & 34   & -0.6$\pm$0.8  & -0.7$\pm$ 0.9 & -0.5$\pm$ 1.2 \\ 
2-3       & 2.34                      & 12   & -1.4$\pm$1.7  & -3.1$\pm$ 3.4 & -1.2$\pm$ 2.0 \\ 
3-4       & 3.36                      &  6   &  1.3$\pm$4.2  &  3.6$\pm$12.2 & -1.6$\pm$ 4.7 \\
4-5       & 4.38                      &  5   &  7.0$\pm$10.1 & 42  $\pm$39   &  5.2$\pm$10.9 \\
\end{tabular} \end{ruledtabular}
\end{table}

\begin{table}
\caption{\label{tbl:anchg}
Charged hadron transverse single-spin asymmetry values and statistical
uncertainties.  
An \AN\ scale uncertainty of $\pm$35\% is not included.
}
\begin{ruledtabular} \begin{tabular}{cccc}
\pperp\   & $\langle \pperp \rangle$ & \ANneg\        & \ANpos\        \\ 
(GeV/$c$) & (GeV/$c$)                & (\%)           & (\%)           \\ 
\hline
0.5-1     & 0.70                     & -0.38$\pm$0.42 & -0.09$\pm$0.41 \\ 
1-2       & 1.32                     & -0.12$\pm$0.82 & -0.54$\pm$0.78 \\
2-5       & 2.56                     & -2.1$ \pm$2.7  & -3.1 $\pm$2.6  \\
\end{tabular} \end{ruledtabular}
\end{table}

In this Letter we have presented the first measurement of transverse-spin asymmetries \AN\ at 
mid-rapidity and high \pperp\ at collider energies and the cross section for inclusive charged 
hadrons at mid-rapidity.  NLO pQCD calculations have been found to reproduce experimental results for 
$\pperp > \sim2$~GeV/$c$ not only for the cross section presented here but also for inclusive neutral pion 
and production, indicating that pQCD can be used to interpret the high-\pperp\ asymmetries.
The transverse SSAs observed for mid-rapidity production of both neutral
pions and charged hadrons are consistent with zero within statistical
errors of a few percent, measured over $0.5 < \pperp < 5$ GeV/$c$.  The result is
consistent with the mid-rapidity results for neutral pions at $\sqrt{s} = 20$~GeV \cite{Adams:1994yu}.  
The present measurement is complementary to that of \cite{Adams:2003fx}.  The large
asymmetries observed in neutral pion production at forward rapidity at
$\sqrt{s} = 200$~GeV \cite{Adams:2003fx} are expected to originate from 
partonic processes involving valence quarks ($x > 0.1$),
whereas the particle production at mid-rapidity presented here is
dominated by gluon-gluon and quark-gluon processes ($x < 0.1$).
Our results are consistent with the pQCD expectation that quark-gluon
correlations are suppressed at high \pperp\ and mid-rapidity \cite{Kane:1978nd,Qiu:1998ia}.  A QCD
analysis of the presented \AN\ may lead to constraints on
gluon-Sivers contributions to observed transverse-spin phenomena.  The
present transverse single-spin asymmetries represent an early measurement in a
rigorous program to study transverse proton spin structure at hard scales
using a pQCD framework at RHIC.


We thank the staff of the Collider-Accelerator Department, Magnet
Division, and Physics Department at BNL and the RHIC polarimetry group for
their vital contributions.  We thank W.~Vogelsang for calculations as well
as numerous useful discussions.  We acknowledge support from the 
Department of Energy and NSF (U.S.A.), 
MEXT and JSPS (Japan), CNPq and FAPESP (Brazil), NSFC (China), 
CNRS-IN2P3 and CEA (France), 
BMBF, DAAD, and AvH (Germany), 
OTKA (Hungary), DAE and DST (India), ISF (Israel), 
KRF and CHEP (Korea), RMIST, RAS, and RMAE (Russia), 
VR and KAW (Sweden), U.S. CRDF for the FSU, 
US-Hungarian NSF-OTKA-MTA, and US-Israel BSF.




\end{document}